\documentclass[aps,twocolumn,superscriptaddress,prl,floatfix]{revtex4}
\usepackage{amssymb}
\usepackage{amsmath}
\usepackage{graphicx}

\def\Gq{e^2/h}

\def\SiOx{\mathrm{SiO}_2}

\def\Vb{V_{\mathrm{B}}}
\def\Vt{V_{\mathrm{T}}}

\def\nm{\mathrm{nm}}
\def\kohm{\mathrm{k}\Omega}

\def\sigmin{\sigma_{\mathrm{min}}}
\def\pn{\emph{p-n }}
\def\pp{\emph{p-p}}
\def\Sxx{S_{xx}}
\def\Rxx{R_{xx}}
\def\Sxy{S_{xy}}
\def\Rxy{R_{xy}}
\def\Rodd{\mathrm{R}_{\mathrm{odd}}}

\begin{document}
\title{Snake States in Graphene p-n Junctions}
\author{J.\ R.\ Williams}
\affiliation{School of Engineering and Applied Sciences, Harvard University, Cambridge, MA 02138, USA}
\author{C.\ M.\ Marcus}
\affiliation{Department of Physics, Harvard University, Cambridge, MA 02138, USA}

\date{\today}

\begin{abstract}
 We investigate transport in locally-gated graphene devices, where carriers are injected and collected along, rather than across, the gate edge. Tuning densities into the \pn regime significantly reduces resistance along the \pn interface, while resistance across the interface increases. This provides an experimental signature of snake states, which zig-zag along the \pn interface and remain stable as applied perpendicular magnetic field approaches zero. Snake states appear as a peak in transverse resistance measured along the \pn interface. The generic role of snake states disordered graphene is also discussed.
\end{abstract}
\maketitle

The low-energy spectrum of graphene, a two-dimensional hexagonal lattice of carbon atoms, yields two gapless modes of massless charge carriers that can be doped either $n$-type (electron-like) or $p$-type (hole-like) by contacts, electrostatic gates, or charged impurities on the graphene surface or within the supporting substrate. The unique band structure of graphene has surprising consequences in transport, including half-integer quantized conductance of $4(n+1/2)$ $\Gq$~ in the quantum Hall regime \cite{Gusynin05, Peres06, Novoselov05, Zhang05}, and finite minimum conductivity $\sigmin \sim \Gq$~\cite{Peres06, Geim07}.  Recently, the ability to locally control density and carrier type by electrostatic gates has resulted in device configurations with adjacent $n$-type and $p$-type regions, separated by an electrically tunable \pn junction (PNJ) \cite{Huard07, Williams07, Ozyilmaz07}. The absence of a gap allows ballistic carriers that approach the PNJ normal to the interface to pass through without reflection, while carriers incident at oblique angles are reflected~\cite{Cheianov06}.  

Besides intentionally gated \pn devices \cite{Huard07, Williams07, Ozyilmaz07}, \pn interfaces play an important role conduction near the charge-neutrality point (CNP) in disordered graphene \cite{Martin07, Adam07, Cheianov07}. Near the CNP, disordered graphene consists of puddles of $p$ and $n$ regions whose boundaries---contours of zero density---form a percolating network of PNJs. Transport modes moving along \pn interfaces contribute to conduction in the percolation region and alter the expected value of conductance, $\sigma_{min}$, at the CNP.

\begin{figure}[b]
\center \label{fig1}
\includegraphics[width=3 in]{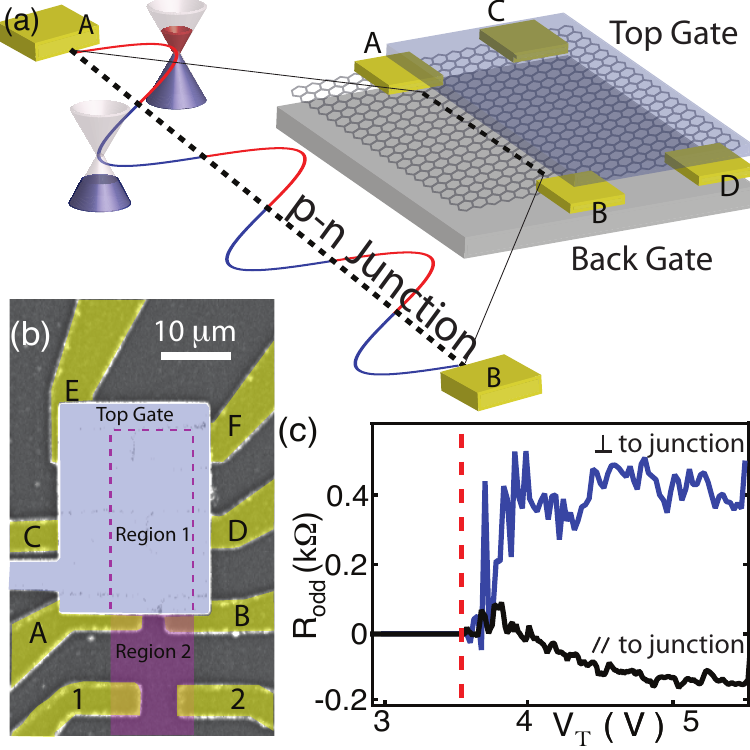}
\caption{\footnotesize{(a) Schematic of the device. Contacts A and B are partially under and partially outside the top gate. Under certain values of $\Vt$ and $\Vb$, a PNJ forms (black dashed line), connecting contacts A and B. A close-up of the PNJ shows the modulation of the density across the junction that allows for an additional conduction channel to appear between contacts A and B in a magnetic field $B$. The change of the Lorentz force (a consequence of the changing sign of the carrier charge) creates a snake-shaped trajectory between the two contacts. (b) False-color image of a device similar to the one studied. (c) The signature of the additional conduction is a reduction of resistance parallel to the junction: for four-terminal resistance measured parallel to the PNJ, $\Rodd$ (defined in the text) is negative (black trace), whereas the four-terminal resistance measured perpendicular to the junction yields a postive $\Rodd$ (blue trace).}}
\end{figure}

In this Letter, we report low-temperature transport in a locally-gated graphene sample with one edge of the gate running between electrical contacts, allowing transport measurements {\em along} the PNJ [see Fig.~1(a)].  Transport is investigated along and away from the gate edge as a function of top gate voltage, $\Vt$, back gate voltage, $\Vb$, and perpendicular magnetic field, $B$. At zero and low magnetic fields, creating a PNJ along the gate edge decreases the longitudinal resistance along the edge and simultaneously increases resistance across the PNJ. These observations indicate the presence of current carrying modes confined to the \pn interface. As the perpendicular field is increased toward the quantum Hall regime, PNJ modes evolve into counter-propagating quantum Hall edge states moving along the \pn interface, with snake states altering transport at the transition between hall plateaus. At low magnetic fields, increasing the electric field across the junction via gate voltages reduces the contribution to conduction of the PNJ interface state, presumably reflecting a destabilization of edge modes with increased potential gradients.

The basic mechanism leading to conduction along the PNJ is easily visualized at small but nonzero perpendicular magnetic field, where a change in sign of charge carriers across the PNJ produces a change in the direction of the Lorentz force, causing classical trajectories to curve back toward the interface from both sides [Fig.~1(a)], similar to so-called ``snake states" that propagate in 2D conductors along contour lines of an inhomogeneous magnetic field~\cite{Muller92}. Snake states in graphene have been considered theoretically for inhomogeneous fields~\cite{Oroszlany08, Ghosh08}, uniform fields at \pn  interfaces~\cite{Beenakker08}, and warped~\cite{Prada10} and folded~\cite{Rainis10} graphene sheets. For the PNJ case~\cite{Beenakker08}, it was shown that classical trajectories similar to that of Fig.~1(a) can exist at the interface of $p$ and $n$ regions and that snake states continue to contribute to transport as $B$ approaches zero. Related guiding of plasmons along PNJ interfaces has also been considered theoretically~\cite{Mishchenko10, Shen10}. 

The main experimental observation of this Letter is illustrated in Fig.~1: while the presence of a PNJ increases resistance across a gate-controlled density interface, resistance along the same interface decreases when a PNJ is formed. Unlike the case of uniformly gated samples, resistances of gated PNJs are typically not symmetric with respect to the charge-neutrality point (CNP), since the top-gate voltage, $\Vt$, controls density only on one side of the junction. Following Ref.~\cite{Stander09}, we extract the contribution to resistance from just the PNJ by separating out the part of the four terminal resistance, R, that is odd relative to the CNP, $\Rodd(\Vt) = \mathrm{R}(\Vt) - \mathrm{R}(2\Vt^{\mathrm{CNP}}-\Vt)$, where $\Vt^{\mathrm{CNP}}$ is the gate voltage that yields charge neutrality under the top gate. When $\Vt$ yields a unipolar sample (with no PNJ), we take $\Rodd(\Vt)  = 0$. The well-known increase in resistance {\it across} a PNJ~\cite{Huard07, Williams07} is observed by sourcing current at contact C with contact 1 grounded, and measuring the voltage between contacts D and 2 [see Fig.~1(b) for geometry]. In Fig.~1(c), a PNJ is present for values of $\Vt$ to the right of the red dashed line, where $\Rodd$ is positive [blue trace in Fig. 1(c)], as expected in this configuration. In contrast, $\Rodd$ is negative [black trace in Fig. 1(c)] for R measured {\it along} the PNJ, with current source at C, ground at D, and voltage measured between A and B. We interpret the negative value of $\Rodd$ in this configuration as the signature of an additional conduction channel at the \pn interface.

\begin{figure*}[t!]
\center \label{fig2}
\includegraphics[width=6 in]{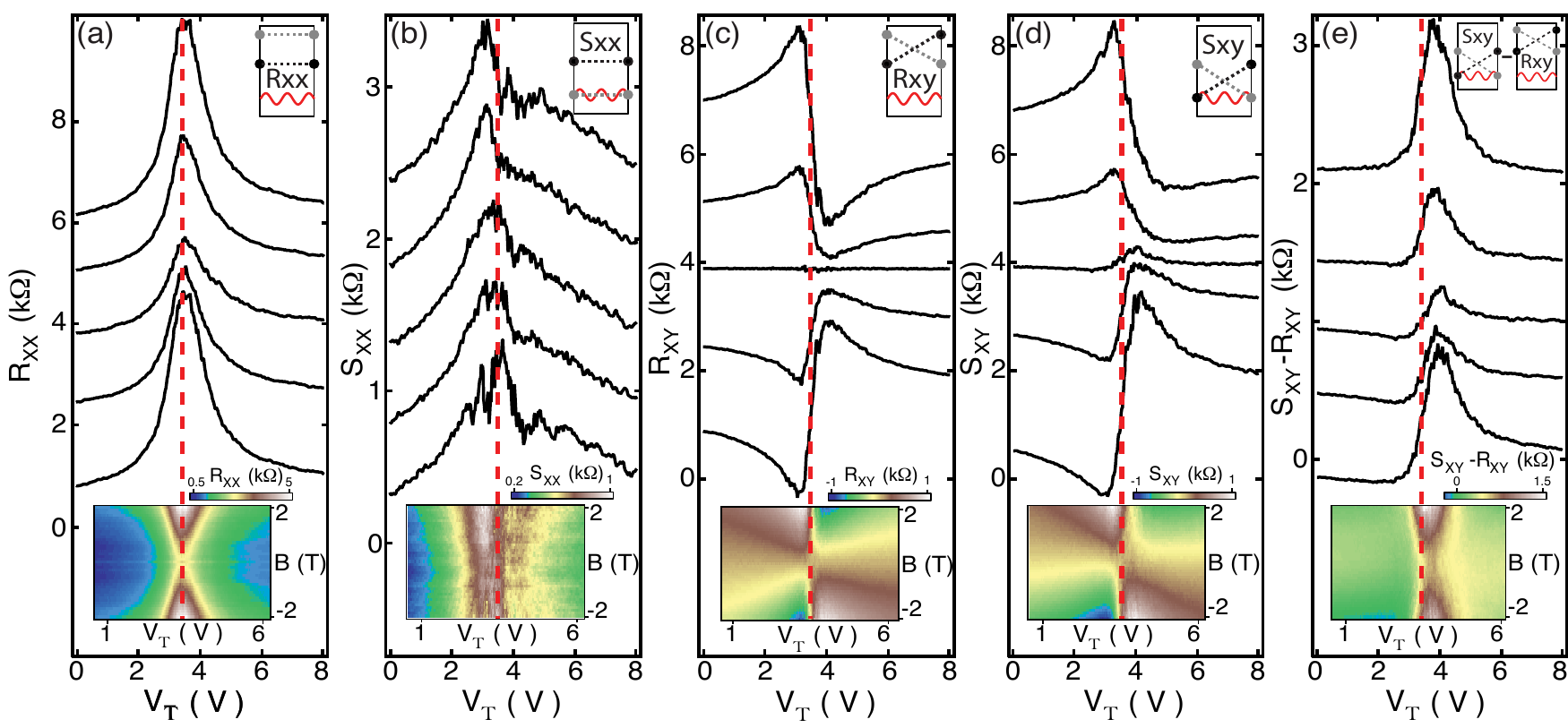}
\caption{\footnotesize{Measurements of the four quantities $\Rxx, \Sxx, \Rxy$ and $\Sxy$ as a function of $\Vt$ (black traces, offset intentional) for $B$ between $\pm$2T in 0.5T steps at a back-gate voltage of -20V. The measurement scheme is shown in the upper right inset (black contacts are the current leads and grey are the voltage leads) in each panel. (a) $\Rxx(\Vt)$. The red dashed line locates the resistance maximum at all $B$ fields and indicates the CNP for Region 1. Lower inset: $\Rxx(\Vt, B)$ demonstrating that the CNP does not change over the $B$ range explored.  (b) $\Sxx(\Vt)$. A drop in resistance of 0.3 - 0.5$\kohm$ is observed at the transition from  $\pp$ regime to $\pn$ regime (red dashed line), indicating that an additional conduction channel has been introduced at the \pn interface. This resistance drop occurs for the entire $B$ field range (lower inset). (c) $\Rxy(\Vt)$. Hall resistance measurement where all the contacts are under the top gate reveals curves that are antisymmetric with respect to the CNP. In contrast, $\Sxy$ (d) has larger resistance on the \pn side of the CNP than $\Rxy$, consistent with an additional conduction channel present at the \pn interface. This additional amount is quantified in (e) where a plot of $\Sxy \mathrm{-} \Rxy$ shows its largest value on the bipolar regime for all $B$ fields (lower inset).}}
\end{figure*}

Graphene flakes made from mechanically exfoliated HOPG were placed on a degenerately doped Si substrate with a 300$\nm$  $\SiOx$ layer.  Potential single layers were identified optically, and electrical contacts (5$\nm$ Ti/40$\nm$ Au) were subsequently patterned using electron-beam lithography and thermal evaporation. An unpatterned 30$\nm$ AlO$_2$ dielectric layer, produced by atomic layer deposition, with an NO$_2$ functionalization layer ~\cite{Williams07}, was deposited on top of the graphene and the contacts. A top gate was patterned as shown in Fig.~1 using an aligned electron beam lithography step. The doped substrate is used as a back gate, controlling density globally; density under the top gate is set by voltage $\Vt$ (Region 1). Contacts straddled by the top gate [labeled A and B in Fig. 1(b)] allow transport measurements along the PNJ. Measurements were made at temperature a of 4K using a 10 nA current-bias measured using a lock-in amplifier. Two similar devices were measured, yielding similar results; data from one device is presented here. Magnetic fields up to 8T were applied perpendicular to the graphene plane. The device reported here shows QH signatures of single-layer graphene, i.e., conductance quantization at 2, 6, 10....
$\Gq$ and has a CNP at $\Vb=40$V. All presented data were taken in the range -40V $<\, \Vb\, <$ 20V (always $p$-type in Region 2, outside the top-gate region). Similar results were obtained (data not shown) for a smaller range of $\Vb>40$V.

In Fig.~2, longitudinal resistance $\Sxx$=$R_{CD,AB}$ and transverse resistance $\Sxy$=$R_{AD,BC}$ along the PNJ are compared to $\Rxx$=$R_{CD,EF}$ and $\Rxy$=$R_{CF,DE}$, with all contacts beneath the top gate.   ($R_{ij,kl}$ denotes a four terminal resistance with current applied at $i$ and $j$ and voltage measured at $k$ and $l$.).  Gate-voltage dependence of $\Rxx$ shows behavior typical of uniformly gated samples~\cite{Novoselov04}: A fairly symmetric peak in $\Rxx$ around $\Vt$$^{\mathrm{CNP}}\sim$ 3.6V identifies the CNP under the gate. For $\Vt$$<\Vt$$^{\mathrm{CNP}}$, the device is unipolar (\pp'); for $\Vt$$>\Vt$$^{\mathrm{CNP}}$, the device is bipolar, with a PNJ along an interface between contacts A and B.  A 2D plot of $\Rxx(\Vt, B)$ [Fig. 2(a), inset] shows that while the resistance increases with increasing $|B|$, $\Vt$$^{\mathrm{CNP}}$ does not depend on $B$. $\Sxx$($\Vt$), measured along the PNJ, differs qualitatively from $\Rxx$ as well as from previously measured longitudinal resistances measured across PNJs~\cite{Huard07, Williams07, Ozyilmaz07}. Notably, in the bipolar regime, $\Sxx$($\Vt$) decreases by $\sim$0.3$\kohm$, producing a trace that is \emph{lower} on the \pn side of the CNP, including at $B=0$ [Fig.~2(b), inset].  

The Hall resistance in the top-gated region, $\Rxy(\Vt)$, is similar to data from single-gate graphene~\cite{Novoselov04} [Fig 2(c)]: As the CNP is crossed, $\Rxy$ changes sign, indicating a change in carrier type ($p\rightarrow n$) as a function of $\Vt$. These curves are antisymmetric with respect to the CNP and $B$.  On the other hand, the Hall resistance that involves the PNJ, $\Sxy$, is not antisymmetric with respect to the CNP; it is larger on the \pn side [Fig. 2(d)]. The difference in Hall resistances, $\Sxy \mathrm{-} \Rxy$ is positive in the \pn regime throughout the measured range of $B$, including zero [Fig.~2(e)].  A measurement similar to $\Sxy$ was previously used to study electron focusing in a 2D electron gas~\cite{vanHouten89}. Here, like in Ref.~\cite{vanHouten89}, an increase in $\Sxy$ is indicative of enhanced transport between contacts A and B, as is observed here. It is important to note that the simultaneous increase in $\Sxy$ and decrease in $\Sxx$ are consistent with enhanced transport along the PNJ and rule out simpler explanations of the phenomena, like a position-dependent CNP within the sheet of graphene. 

\begin{figure}
\center \label{fig3}
\includegraphics[width=3 in]{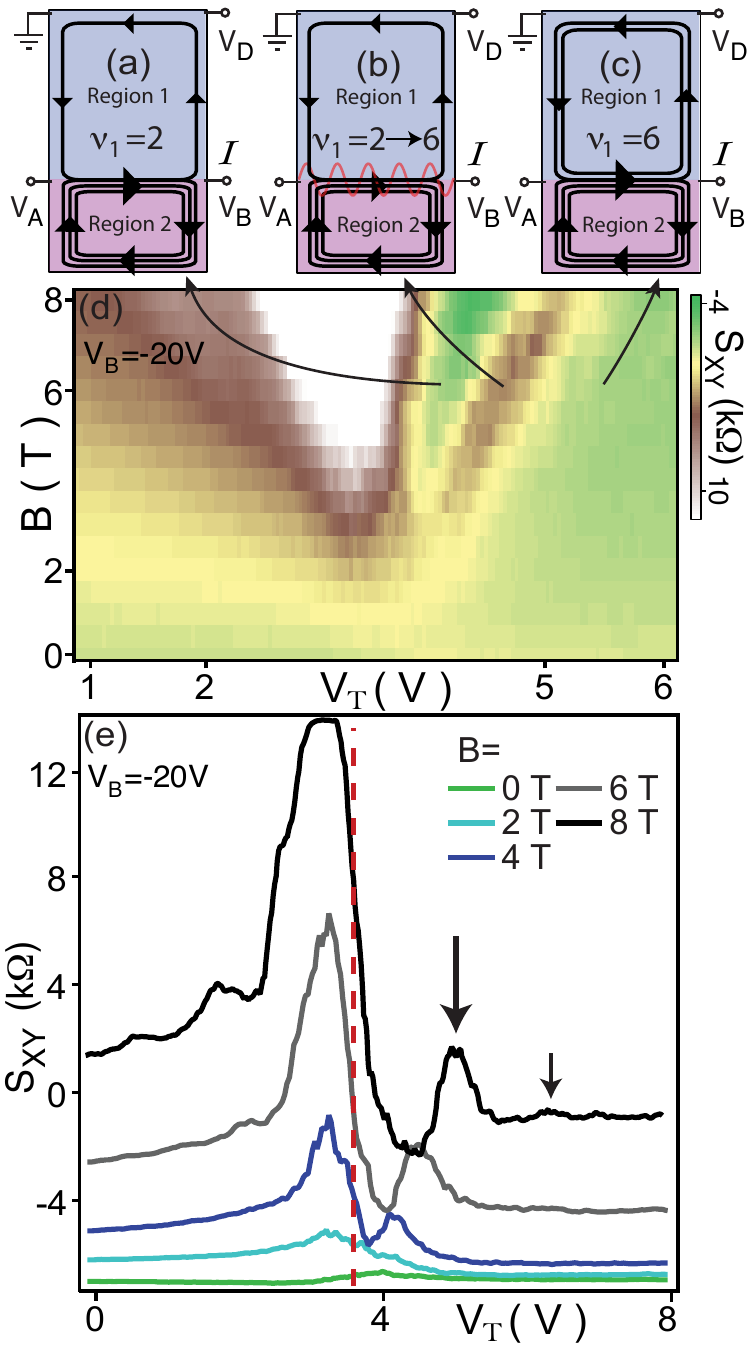}
\caption{\footnotesize{(a-c) Schematic of transport in the quantum Hall regime for a constant $\nu_{R2}$=-10, while $\nu_{R1}$ progress from 2 (a), transition from 2 to 6 (b), and 6 (c). At the transition between filling factors 2 and 6 (b), snake states enhance $\Sxy$ in a manner similar to Fig. 2, with an amplitude that increases as B increases. (d) $\Sxy(\Vt, B)$ reveals expected plateaus of 1/10, 1/6 and 1/2 $h/e^2$in the unipolar regime ($\Vt<$ 4V) and deviations from these values in the bipolar regime. An enhanced value of $\Sxy$ is observed at the transition between $\nu_{R1}$ = 2 and 6 and is attributed to the presence of snake states at the $\pn$ interface. (e) Constant-B-field cuts of (d) for B between 0 and 8T in 2T steps. Cuts from 0 to 6T are shifted downward from the B=8T cut for clarity. }}
\end{figure}

In the quantum Hall regime ($B> 4$T), snake states also alter the nature of transport along a PNJ. $\Sxy$ in the unipolar regime, with the filling factor in Region 2 = -10, shows typical quantum Hall behavior for graphene, with plateaus at $1/2$, $1/6$ and $1/10$ in units of $h/e^2$, and depends only on the filling factor of Region 1,  $\nu_{R1}$ [black curve in Fig. 3(e), to the left of the red dashed line]. In the presence of a PNJ, the value of $\Sxy$ deviates from the typical values for graphene. When Region 1 is on a Hall plateau, i.e. for configurations shown schematically in Figs. 3(a,c), edge states equilibrate at contact A and are all ejected at the same voltage towards contact B. The values for $\Sxy$ are a result of edge state propagation and can be calculated via the Landauer-Buttiker formula, resulting in a close match with the experimental values of 2.3k$\Omega$ and 1.6k$\Omega$ for $\nu_{R1}$=2 and 6, respectively (see Supplementary Info). In addition to the modified $\Sxy$ resistances, a peak in resistance appears at $\Vt$=5.5V at 8T [Fig. 3(e), indicated by black arrow] of magnitude $\sim$2$\kohm$. A smaller, but evident, peak is also visible at $\Vt$=6.5 V [smaller black arrow, Fig. 3(e)]. A plot $\Sxy(\Vt, B)$ shows that the stronger peak (large black arrow) moves linearly away from the CNP of Region 1 as $B$ is increased [Fig. 3(d)] and follows the transition $\nu_{R1}$=2$\rightarrow$6, suggesting that the peak is due to contributions to the resistance from $\rho_{xx}$. This case is unlikely as the contribution from $\rho_{xx}$ in the \pp ~regime at the transition $\nu_{R1}$=-6$\rightarrow$-2 is not nearly as prevalent as this peak observed in the \pn regime.  In between Landau levels, transport in Region 1 is allowed to occur in the bulk, as is the case near B=0. In the same way that $\Sxy$ is enhanced at low fields in Fig. 2(d), it is enhanced here at the transition from $\nu_{R1}$=2$\rightarrow$6 by the presence of a snake state at the interface [shown schematically in Fig. 3(b)]. Fig. 3(e) shows one-dimensional cuts of Fig. 3(d) starting at B=0 to B=8T in 2T steps, with each B-value curve shifted downward from the B=8T value for clarity. Here it is evident that the enhancement  of $\Sxy$ at B=0 gradually evolves into the peak in the QH regime, suggesting that the low- and high-field phenomena have similar origins. Further, confinement at the interface of the PNJ should be enhanced by a magnetic field, resulting in a stronger peak in $\Sxy$, as is also evident in Fig. 3(e).

Increasing $\Vb$ produces a voltage drop across the PNJ [$\Delta$V=V(Region 1)-V(Region 2), where both voltages are measured at the position of the peak in $\Sxy \mathrm{-} \Rxy$] that increases, allowing for the transverse-electric-field dependence of the snake state to be investigated. The $\Delta$V dependence of the peak resistance in $\Sxy \mathrm{-} \Rxy$ [from Fig. 2(e)] is shown in Fig. 4, plotted for $B$ between 0 and 2T, in 0.5T increments in the $\Delta$V range of $\sim$23V to 83V (black lines are guides to the eye). It is found that the position of the peak moves linearly in the ($\Vt$,$\Vb$) space (data not shown) but decreases as the magnitude of $\Delta$V is increased, i.e. as the electric field perpendicular to the junction increases.   The change in resistance gets stronger for increases in the $B$ field, changing $\sim$0.2$\kohm$ at $B$=0T to $\sim$1$\kohm$ at $B$=2T over the $\Delta$V range shown here.

\begin{figure}
\center \label{fig4}
\includegraphics[width=3 in]{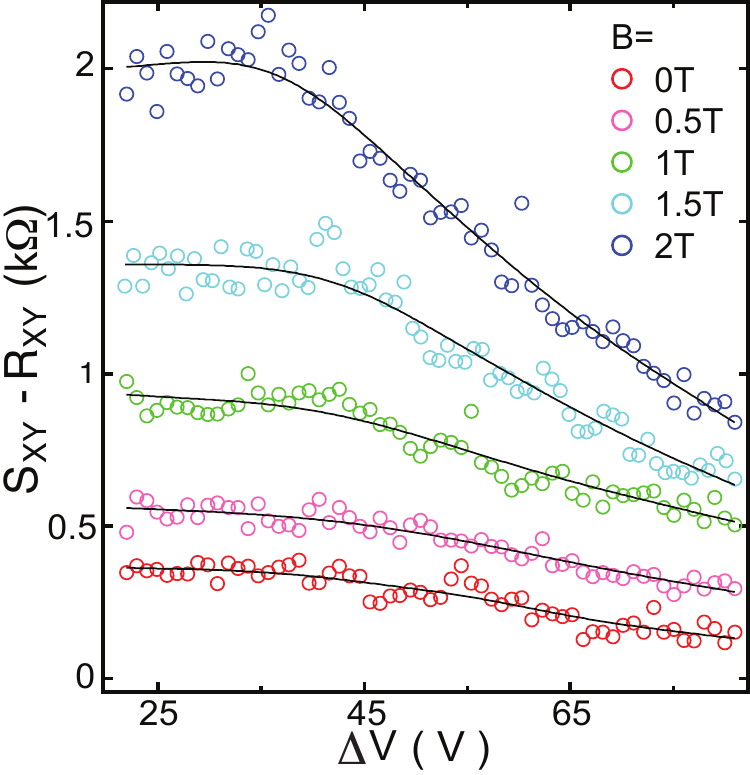}
\caption{\footnotesize{$\Delta$V dependence of the difference $\Sxy \mathrm{-} \Rxy$ for $B$ between 0 and 2T in 0.5T increments. $\Sxy \mathrm{-} \Rxy$ is reduced as $\Delta$V becomes larger for all $B$ fields. Solid black lines are guides to the eye. The decrease in resistance in this $\Delta$V range increases as the perpendicular field is increased, rising from $\sim$0.2$\kohm$ at $B$=0T to $\sim$1$\kohm$ at $B$=2T.}}
\end{figure}

An additional channel of conductance along the PNJ would alter the picture of charge transport at the charge-neutrality point, changing the value of $\sigmin$, the conductance at this point. Charge transport in disordered graphene samples has been studied experimentally~\cite{Novoselov05, Tan07}  and theoretical predictions have been made for $\sigmin$~\cite{Rossi08, Fogler08}, however consensus has yet to be reached. Taking into account the resistance of the PNJs and appropriate values for the size of density fluctuations, a value $\sigmin$$\sim$2.5$\Gq$ was obtained, 2 to 6 times lower than the experimentally reported values~\cite{Novoselov05, Tan07}.  The additional conduction along the interface could be a source conductance that brings this theoretical value closer to the experimental values. 

We thank L.~S. Levitov and D.~A. Abanin for useful
discussions. Research supported in part by INDEX, an NRI Center, and Harvard NSEC.

\end{document}